\pdfoutput=1
\documentclass[conference]{IEEEtran}
\IEEEoverridecommandlockouts
\PassOptionsToPackage{table, x11names, dvipsnames, svgnames}{xcolor}
\usepackage{lipsum}
\usepackage{ dsfont }
\usepackage{cite}
\usepackage{bbm}
\usepackage{amsmath,amssymb,amsfonts}
\usepackage{algorithm, algorithmic}
\usepackage{graphicx}
\usepackage{textcomp}
\usepackage{xcolor}
\usepackage[english]{babel}
\usepackage[acronym,nonumberlist]{glossaries}
\usepackage{glossaries-prefix}
\usepackage{pgfplots}
\usepgfplotslibrary{fillbetween}
\usepackage{amssymb}
\usepackage{import}
\usepackage{bm}
\usepackage{siunitx}
\usepackage{subcaption}
\usepackage{braket}

\usepackage{hyperref}
\hypersetup{%
linktocpage=true, 
colorlinks=false,
pdfborder={0 0 0}, 
breaklinks=true, pdfpagemode=UseNone, pageanchor=true, pdfpagemode=UseOutlines,%
plainpages=false, bookmarksnumbered, bookmarksopen=true, bookmarksopenlevel=1,%
hypertexnames=true, pdfhighlight=/O,
pdftitle={Batch Quantum Reinforcement Learning},%
pdfauthor={Maniraman Periyasam},%
pdfsubject={QML},%
pdfkeywords={},%
pdfcreator={pdfLaTeX},%
pdfproducer={LaTeX with hyperref}%
}

\usepackage{pgfplots}
\pgfplotsset{width=7cm,compat=1.8}

\usepackage{multirow}
\usepackage{color}
\usepackage{tabularray}

\usepackage[inline]{enumitem}

\usepackage[capitalise]{cleveref}
\usepackage{braket}
\usepackage{multirow, makecell}
\usepackage{placeins}

\newacronym[shortplural=GMMs]{GMM}{GMM}{Gaussian mixture model}
\newacronym[shortplural=HMMs]{HMM}{HMM}{hidden Markov model}
\newacronym[shortplural=DNNs]{DNN}{DNN}{deep neural network}
\newacronym[shortplural=SVDs]{SVD}{SVD}{singular value decomposition}
\newacronym[
    prefixfirst={a\ },
    prefix={an\ }
]{MCTS}{MCTS}{Monte Carlo tree search}
\newacronym[prefixfirst={a\ },prefix={an\ }]{MDP}{MDP}{Markov decision process}
\newacronym{CMDP}{CMDP}{constrained Markov decision process}
\newacronym{DDPG}{DDPG}{deep deterministic policy gradient}
\newacronym{RL}{RL}{reinforcement learning}
\newacronym{DRL}{DRL}{deep reinforcement learning}
\newacronym[shortplural=DTs]{DT}{DT}{decision tree}
\newacronym{SMT}{SMT}{satisfiability modulo theories}
\newacronym{IL}{IL}{Imitation Learning}
\newacronym[shortplural=CNNs]{CNN}{CNN}{convolutional neural network}
\newacronym[shortplural=DQNs]{DQN}{DQN}{deep Q-networks}
\newacronym{AI}{AI}{artificial intelligence}
\newacronym{PPO}{PPO}{proximal policy optimization}
\newacronym{ML}{ML}{machine learning}
\newacronym{QML}{QML}{quantum machine learning}
\newacronym{NISQ}{NISQ}{noisy intermediate scale quantum}
\newacronym{QC}{QC}{quantum circuit}
\newacronym{VQC}{VQC}{variational quantum circuit}
\newacronym{VQ-DQN}{VQ-DQN}{variational quantum deep Q-networks }
\newacronym{MNIST}{MNIST}{modified national institute of standards and technology}
\newacronym{FIM}{FIM}{Fisher information matrix}
\newacronym{IDU}{IDU}{incremental data-uploading}
\newacronym{DRU}{DRU}{data re-uploading}
\newacronym{QRL}{QRL}{quantum reinforcement learning}
\newacronym{BCQ}{BCQ}{batch-constraint deep Q-Learning}
\newacronym{SPSA}{SPSA}{simultaneous perturbation stochastic approximation}
\newacronym{BCQQ}{BCQQ}{batch-constraint quantum Q-learning}

\usepackage{lineno,todonotes}

\usepackage{yquant}
\usepackage{braket}
\usetikzlibrary{quotes, fit}
\usepackage{environ}

\makeatletter
\newsavebox{\measure@tikzpicture}
\NewEnviron{scaletikzpicturetowidth}[1]{%
  \def\tikz@width{#1}%
  \begin{lrbox}{\measure@tikzpicture}%
  \BODY
  \end{lrbox}%
  \pgfmathparse{#1/\wd\measure@tikzpicture}%
  \BODY
}
\makeatother

\newcommand\copyrighttext{%
  \footnotesize \textcopyright This work has been submitted to the IEEE for possible publication. Copyright may be transferred without notice, after which this version may no longer be accessible}
\newcommand\copyrightnotice{%
\begin{tikzpicture}[remember picture,overlay]
\node[anchor=south,yshift=10pt] at (current page.south) {\fbox{\parbox{\dimexpr\textwidth-\fboxsep-\fboxrule\relax}{\copyrighttext}}};
\end{tikzpicture}%
}

\begin{document}



\title{BCQQ: Batch-Constraint Quantum Q-Learning with Cyclic Data Re-uploading
}

\author{
%
\IEEEauthorblockN{Maniraman Periyasamy, Marc Hölle, Marco Wiedmann, Daniel D. Scherer, Axel Plinge, Christopher Mutschler}
\IEEEauthorblockA{\textit{Fraunhofer IIS, Fraunhofer Institute for Integrated Circuits IIS},
Nuremberg, Germany \\\vspace{1mm}}
}

\maketitle
\copyrightnotice

\begin{abstract} 
Deep reinforcement learning (DRL) often requires a large number of data and environment interactions, making the training process time-consuming. This challenge is further exacerbated in the case of batch RL, where the agent is trained solely on a pre-collected dataset without environment interactions. Recent advancements in quantum computing suggest that quantum models might require less data for training compared to classical methods. In this paper, we investigate this potential advantage by proposing a batch RL algorithm that utilizes \glspl{VQC} as function approximators within the discrete batch-constraint deep Q-learning (BCQ) algorithm. Additionally, we introduce a novel data re-uploading scheme by cyclically shifting the order of input variables in the data encoding layers. We evaluate the efficiency of our algorithm on the OpenAI CartPole environment and compare its performance to the classical neural network-based discrete BCQ.
\end{abstract}


\glsresetall
\section{Introduction}
\label{sec:introduction}
The challenge of applying \gls{RL} in real-world problems lies in its training process. In contrast to fully data-driven \gls{ML} procedures such as supervised learning, RL learns via environment interactions. An RL agent follows a policy and chooses actions that change the state of the environment, for which it receives a reward (possibly after each interaction). The agent's objective is to learn a policy that maximizes the long-term reward. Using this framework, agents based on \glspl{DNN} have been remarkably successful in a variety of complex tasks, including super-human performance in computationally-hard board games \cite{Silver2018} and discovering fast matrix multiplication algorithms \cite{Fawzi2022}.

Unfortunately, this interactive approach is not feasible in many safety-critical scenarios, that potentially benefit from RL, e.g., robotics or healthcare. While in general it is possible to train RL agents in a simulator, it is often non-trivial to deploy them in the real world due to the domain gap between simulation and reality. In these cases, it would be beneficial to utilize real-world data, gathered by an expert operator, and train in a purely data-driven, offline fashion. 
However, current offline algorithms require large datasets to match the performance of algorithms trained with environment interactions \cite{Agarwal2020}. This is an impediment in domains where only limited data is available. One of the main reasons for this performance loss is that the distribution of states and actions during testing can drastically differ from the training data. Although RL research heavily tackles this lately, these challenges still remain mostly unresolved.

With the advent of the first practical quantum computers, so-called \gls{NISQ} devices, it is natural to investigate whether this new computing paradigm could be leveraged to improve RL.
Factors like low gate fidelity and coherence times of these \gls{NISQ} devices cause applied research to focus on hybrid quantum-classical schemes such as \glspl{VQC} \cite{Kiss2022, OMalley2016, Yun2023}. These can be considered as the quantum analog to classical \glspl{DNN} and are therefore also called quantum neural networks. Theoretical work indicates that VQCs are more data-efficient than classical \gls{ML} methods ~\cite{caro2022generalization}. As data can become a bottleneck when learning offline, we are interested in investigating if this theoretical advantage can be turned into a practical performance gain. This would translate to a \gls{QRL} algorithm \cite{Meyer2022} that can learn a policy from a small dataset and outperform a classical policy trained on the same data.
Currently, the limited number of qubits together with the corresponding hardware topology and the complexity of numerical simulations make it intractable to benchmark QRL algorithms on state-of-the-art environments on a large scale. However, proof-of-concept experiments can be executed in low-complexity environments such as OpenAI's CartPole~\cite{Brockman2016}.

Our contribution in this paper is two-fold. First, we show how to apply function approximation within the discrete batch-constraint deep Q-learning algorithm~\cite{Fujimoto2019discrete} with VQCs, where we find a performance advantage over the classical counterpart in CartPole. Near-optimal performance can be achieved in a low data regime, where a classical agent with similar number of parameters that is able to solve CartPole in an online setting, fails offline. Second, we present a cyclic data re-uploading scheme that proved to be advantageous in this batch RL context and analyze the effective dimension of the resulting \gls{VQC}.
\section{Theoretical Background}
\label{sec:background}

\subsection{General Framework of Reinforcement Learning}
On an abstract level, RL can be modelled in the framework of Markov-Decision-Problems (MDPs) \cite{van2012reinforcement}, in which an agent interacts with its environment. An MDP is defined by a set of states \(\mathcal{S}\), actions \(\mathcal{A}\), reward function \(R: \mathcal{S} \times \mathcal{A} \rightarrow \mathds{R}\) and discount factor \(\gamma \in [0, 1]\).
The goal is to find an optimal policy, that assigns each action \(a \in \mathcal{A}\) a probability \(\pi(a|s)\) with which the agent should take that particular action, given that the environment is in state \(s \in \mathcal{S}\). In this case, optimality means that the policy should maximize the expected, discounted reward for any given initial state \(s_0 \in \mathcal{S}\), i.e.
\begin{equation}
    \pi^* \in \underset{\pi}{\mathrm{argmax}} \sum_{t=0}^{\infty} \underset{\substack{ a_t \sim \pi \\ s_{t+1} \sim T}}{\mathds{E}}\gamma^t R(s_t, a_t).
\end{equation}

One usually defines a \textit{state-action value function} or \textit{Q-function}
\begin{equation}
    Q_\pi(s, a) = \underset{s_0 \sim T}{\mathds{E}} \left(R(a, s) + \gamma \sum_{t=0}^{\infty} \underset{\substack{ a_t \sim \pi \\ s_{t+1} \sim T}}{\mathds{E}}\gamma^t R(s_t, a_t)\right) 
\end{equation}
as the expected, discounted cumulative reward from choosing action \(a\) in state \(s\) and following the policy \(\pi\) from there on. It can be shown that the Q-function of an optimal policy must satisfy a recursive relation known as the Bellman optimality equation \cite{sutton2018reinforcement}
\begin{equation}
    \label{eq:Bellmann}
    Q^*(s, a) = \underset{s' \sim T}{\mathds{E}} \left(R(s, a) + \gamma \underset{a'\in \mathcal{A}}{\text{max}} Q^*(s', a')\right).
\end{equation}
On the other hand, given the optimal Q-function \(Q^*\) one can represent the corresponding optimal policy as
\begin{equation}
\label{eq:InducedPolicy}
    \pi^*(a|s) = \frac{1}{\mathcal{N}}\begin{cases}1\text{ , if } a \in \underset{a' \in \mathcal{A}}{\text{argmax }} Q^*(s, a') \\0 \text{ , else} \end{cases}
\end{equation}
where \(\mathcal{N}\) is a normalization factor assuring that all probabilities sum to unity.
\subsection{Online vs. Batch and On-Policy vs. Off-Policy RL}
The task of learning an optimal policy can be approached in different ways. One fundamental distinction between \gls{RL} algorithms is the kind of data used in the training process. If the learning algorithm interacts with the environment during the training process, it is called an \textit{online} algorithm. 
However, this does not mean that the algorithm needs to be \textit{on-policy}, i.e. exclusively use it's current estimate of an optimal policy to gather data. \textit{Off-policy} algorithms on the other hand often employ explorative policies to gather experience from the environment and train a separate policy that should eventually solve the given task.
In contrast to this, \textit{batch} or \textit{offline} algorithms only need access to data that was collected from the environment beforehand. In principle, any off-policy learning algorithm can be used for batch RL, but it requires special care to avoid problems during training \cite{Fujimoto2019discrete}. Many offline RL approaches rely on Q-learning with \gls{DQN}, as explained in the next chapter.

\subsection{Deep Q-Learning}
Q-learning relies on the fact that an optimal policy is induced by a Q-function that solves the Bellman equation \eqref{eq:Bellmann}.
A popular approach for large-scale problems is \gls{DQN}, which uses \glspl{DNN} as parametrized function approximators for \(Q^*\) \cite{Mnih2015}. They can be trained on a loss function that is derived from the Bellman equation \eqref{eq:Bellmann}.

\begin{equation}
    l(\theta) = \underset{M}{\mathds{E}}\left(r + \gamma \underset{a' \in \mathcal{A}}{\text{max}} Q_{\theta'}(s', a') - Q_\theta(s, a)\right)^2
\end{equation}

where, $M$ is the mini-batch sampled from buffer $\mathcal{B}$ consisting of transitions $(s, a, r, s')$.

\subsection{Variational Quantum Circuits for RL}
\label{sec:VQC}
\glspl{VQC} have gained a lot of attention from the quantum computing community in the recent years due to their \gls{NISQ} feasibility. It has already been established, that they are a potentially powerful platform for quantum-enhanced \gls{ML} \cite{schuld2020circuit, chen2020hybrid, blance2021quantum}. While any quantum computation can be decomposed into a sequence of quantum gates, the power of \glspl{VQC} arises from the fact that some of these gates are parameterized by a continuous variable. For example, any single qubit gate can be expressed as a rotation in 3D space acting on the Bloch vector and is therefore parameterized by the three Euler angles corresponding to this rotation. This allows us to use quantum circuits as parameterized function approximators. For a detailed introduction into the framework and theory of quantum computing, we refer the reader to \cite{nielsen2010quantum}.

A VQC that represents a function \(f_\theta(x)\) is composed of three basic components \cite{Mitarai2018}:
\begin{enumerate*}
    \item The data encoding. The unitary \(U(x)\) encodes the input data \(x\) into a quantum state \(\ket{\psi(x)} = U(x)\ket{0}\).
    \item The variational layers. A different unitary \(U(\theta)\) maps the input state \(\ket{\psi(x)}\) onto the output state \(\ket{\phi(x, \theta)} = U(\theta)\ket{\psi(x)}\). A common approach is to decompose \(U(\theta)\) into a set of repeated layers, which contain both parameterized rotations and entangling gates.
    \item The measurement. An observable \(O\) is chosen and its expectation value is estimated over several runs of the circuit. The result represents the output of the VQC 
\end{enumerate*}
\begin{equation}
    \label{eq:VQCExpectationValue}
        f_\theta(x) = \bra{\phi(x, \theta)} O \ket{\phi(x, \theta)}.
    \end{equation}
It has been shown in \cite{Schuld2021, Gil2020} that a given VQC actually realizes a specific Fourier sum
\begin{equation}
    f_\theta(x) = \sum_{\omega \in \Omega} c_\omega(\theta) e^{i\omega x},
\end{equation}
where the available frequency spectrum \(\Omega\) is determined by the data encoding. Repeating the data encoding, i.e. the action of the unitary \(U(x)\), multiple times throughout the circuit (cf. section \nameref{sec:dru}) enlarges the available spectrum. Thus, a broader class of functions can be accessed through this so-called data re-uploading.

To take this analogy even further: Just like NNs, VQCs can be trained from a set of samples \(\{(x, f(x))\}\) of the desired function. The optimization of the parameters \(\theta\) in the training phase of the VQC is carried out on a classical computer, which may evaluate the VQC as a quantum subroutine.
The hope is that VQCs are able to access a broader class of functions with fewer parameters and enable more data efficient learning. Since the full action of the unitaries \(U(x)\) and \(U(\theta)\) is believed to be hard to simulate classically for appropriately designed VQCs, quantum computers might provide an advantage in machine learning tasks.

In the context of QRL, a common approach is to employ VQCs instead of NNs to represent the Q-function in DQN. The state of the environment is used as the input to the VQC. Each available action \(a \in \mathcal{A}\) is assigned an observable \(O_a\), such that the Q-function estimate is given by
\begin{equation}
    Q_\theta(a|s) = \bra{\phi(s, \theta)} O_a \ket{\phi(s, \theta)}
\end{equation}
However, since the output of the VQC is the quantum mechanical expectation value of an observable (ref. equation \eqref{eq:VQCExpectationValue}), the range of possible output values is quite limited. Popular choices for the observables \(O_a\) are combinations of single-qubit Pauli matrices. This restricts the output of the VQCs to the interval \([-1, 1]\). However, the true Q-function might be beyond this range. To address this, one can simply scale each expectation value by a classical weight \(w_a\), which is also inferred from the training process \cite{Skolik2022}.

\subsection{Efficient Gradient Estimation on Quantum Devices}

Since computing an explicit representation of the unitary \(U(\theta)\) of the VQC is classically hard, one cannot efficiently compute the gradient of the VQC output with respect to the parameters \(\theta\) with classical techniques. Although there is a way of obtaining the gradient directly from the quantum device with the \textit{parameter-shift rule} \cite{Schuld19}, this approach still suffers from the fact that it requires \(2p\) expectation value estimations for a circuit with \(p\) parameters, which quickly becomes intractable even for medium sized circuits due to the large number of required circuit runs.
This is why gradient-free optimization schemes, like \gls{SPSA}, which always uses only two calls to the quantum subroutine for each update step, have become of interest to the QML community. It is especially well suited for noisy objective functions which arise when training \glspl{VQC} \cite{pellow2021comparison, Bonet2021, Mihalikova2022}. It computes an approximate gradient, that can be fed to state-of-the-art gradient based optimizers like AMSGrad \cite{reddi2019convergence} to efficiently train medium-sized VQCs \cite{Wiedmann2023}. However, we suspect that this method introduces instabilities in the training of offline \gls{RL} agents, when no appropriate early-stopping criterion is available.

\section{Related Work}
\label{sec:related_work}
The following section starts with a short summary of \gls{QRL}, followed by a survey of relevant batch \gls{RL} literature.

\subsection{Quantum Reinforcement Learning}

The use of quantum computing for \gls{RL} is an emerging field. Dong et al.~\cite{Dong2005, Dong2008} were among its pioneers, proposing an algorithm that learns a state-value function by using the Grover search algorithm \cite{Grover1996}. More recently, Chen et al. \cite{Chen2020} proposed one of the first \gls{VQC}-based \gls{QRL} algorithms, by replacing neural networks in the \gls{DQN} algorithm with \glspl{VQC}. For a detailed picture of \gls{QRL} we refer to Meyer et al. \cite{Meyer2022}. Upon completion of our work, we came across the works of Cheng et al. \cite{Cheng2023offline}, which uses \glspl{VQC} in conservative Q-learning for offline \gls{RL}. There the authors showed that a quantum agent can learn to solve an environment in an offline fashion. However, the authors did not investigate the performance of quantum agents on noisy data or partial trajectories.

\subsection{Batch Reinforcement Learning} 
When \gls{RL} is utilized in scenarios where humans or equipment can be harmed, the agent is usually trained in a simulated environment beforehand. Consequently, the agent's performance might suffer from limited modelling of real-world processes and creating these simulations causes an additional overhead. To resolve this, the objective of batch RL is to learn an optimal policy from a set of real-world data. This training data is gathered by a behavior policy (e.g. expert human operator) interacting with the environment. As an extreme case, one can even consider a behavior policy selecting actions at random. Off-policy algorithms such as \gls{DQN} are similar in spirit to offline learning, since they are in principle agnostic to how the experience was gathered. However, Fujimoto et al. \cite{Fujimoto2019} found that Q-value estimates of \gls{DQN} diverge in an offline setting, leading to the offline trained algorithms performing worse than the same algorithms trained on the same dataset in an online manner. Agarwal et al. \cite{Agarwal2020} demonstrated, that offline trained agents can reach comparable performance on the OpenAI Atari 2600 games \cite{Brockman2016}, but a very large (50 million environment interactions) and diverse dataset was used.
This can be explained by the fact, that algorithms like \gls{DQN} employ a replay buffer of past environment interactions, which causes the gathered data to be correlated to the current policy \cite{Mnih2015}. In offline RL, this correlation is not present and there is a distributional shift between training and testing. Furthermore, out-of-distribution states may arise, if during testing the agent encounters a state that was not part of its training data.

Additionally, Q-learning-based algorithms like \gls{DQN} tend to suffer from overestimating Q-values, due to the objective of maximizing the expected return \cite{Thrun1993}. In the online scenario, this is not as severe, due to corrective feedback from the environment during training. However, since this feedback is missing in an offline setting, this overestimation bias together with the distributional shift causes the policy to extrapolate poorly to unfamiliar states. A problem to which Fujimoto et al. \cite{Fujimoto2019} refer to as extrapolation error. Approaches trying to alleviate this error typically introduce some constraint on the policy to keep it close to the behavior policy. Typically, closeness is determined either directly with respect to a probability metric or penalty terms in the policy update \cite{Levine2020}.

\subsection{Batch-Constraint Deep Q-Learning}
\label{sec:BCQ}

An algorithm implicitly following the first approach is \gls{BCQ} \cite{Fujimoto2019}. The key idea is that in order to avoid the distributional shift, a trained policy should induce a similar state-action visitation to what is observed in the batch. In the following, such policies are called \textit{batch-constrained}. To achieve this, BCQ uses a generative model \(G_{\omega}\) to preselect likely actions according to the batch. The policy is only allowed to choose from this preselection. In the following, we will restrict the discussion of BCQ to the discrete action setting. In this case, the generative model can be understood as a map \(G_{\omega}: \mathcal{S} \rightarrow \Delta \left(\mathcal{A}\right)\) that takes the current environment state as input and outputs the probability with which each action would occur in the batch. In particular, if the batch is filled using transitions from a policy \(\pi_b\) then the generative model should reconstruct this policy, i.e. \(G_\omega(a|s) \approx \pi_b(a|s)\). From this, one can preselect the actions by discarding actions whose probability relative to the most likely one is below a threshold \(\tau\)
\begin{equation}
\label{eq:BatchConstraintActions}
    \tilde{\mathcal{A}}(s) = \left\{a \in \mathcal{A} \Bigg{|} \frac{G_\omega(a|s)}{\underset{\hat{a} \in \mathcal{A}}{\text{max}} G_\omega(\hat{a}|s)} > \tau\right\}.
\end{equation}
Both the policy
\begin{equation*}
    \pi_\theta(a|s) = \frac{1}{\mathcal{N}}\begin{cases}1\text{ , if } a \in \underset{a' \in \tilde{\mathcal{A}}(s)}{\text{argmax}} Q_\theta(s, a') \\0 \text{ , else} \end{cases}
\end{equation*}
and the target in the loss function
\begin{equation*}
    l(\theta) = \underset{(s, a, r, s') \in \mathcal{B}}{\mathds{E}}\left(r + \gamma \underset{a' \in \tilde{\mathcal{A}}(s')}{\text{max}} Q_{\theta'}(s', a') - Q_\theta(s, a)\right)^2 \\
\end{equation*}
are updated to only consider this preselection of actions. The generative model itself is trained with a standard cross-entropy loss
\begin{equation*}
    l(\omega) = - \sum_{(s, a) \in \mathcal{B}}\text{log}\left(G_\omega (a|s)\right).
\end{equation*}

Additionally, to address the overestimation bias of Q-learning towards underrepresented transitions, a technique called \textit{Double DQN} \cite{van2016deep} is employed. Instead of selecting the maximal action with respect to the target network in the Q-learning target, the maximal action with respect to the current Q-network is chosen, but it is still evaluated using the target network. The corresponding loss function is
\begin{align}
    &l(\theta) = \underset{(s, a, r, s') \in \mathcal{B}}{\mathds{E}}\left(r + \gamma Q_{\theta'}(s', a') - Q_\theta(s, a)\right)^2 \\
\text{where} \nonumber \\
    & a' \in \underset{\tilde{a} \in \tilde{\mathcal{A}}(s')}{\text{argmax}} Q_\theta(s', \tilde{a}). \nonumber
\end{align}

In this work, we apply the \gls{VQ-DQN} proposed by \cite{Franz2022} to get an offline \gls{QRL} algorithm which we call \gls{BCQQ}.
\section{BCQQ}
\label{sec:experimental_setup}

Classical batch RL methods like discrete BCQ often struggle to learn an optimal policy in scenarios where high-quality training samples are unavailable in abundance. Loosely interpreted, this behavior suggests that function approximation via classical neural networks requires a large amount of data to effectively approximate a policy. However, the VQCs have shown some indication of approximating a function from far fewer samples in the supervised learning context \cite{caro2022generalization}. 
Therefore, we constructed and conducted a series of experiments to study the capabilities of a VQCs in learning a policy in the batch RL setup.

\subsection{RL Environment and Offline Data Collection}
\label{sec:dataset}
The capabilities of a VQC in learning an optimal policy in an online fashion to solve CartPole environments have already been studied in various instances \cite{Skolik2022, Franz2022}. Therefore, we chose the CartPole-v1 environment from the OpenAI gym \cite{Brockman2016} as the target environment for all the performed experiments. The offline datasets solving real-world problems often do not have trajectories from the optimal policy for a given setup. To mimic the worst-case scenario, we experimented with the most adverse situation, where the data buffer contains trajectories only from a random policy interacting with the CartPole-v1 environment. Buffers with $10^2$, $10^4$, and $10^6$ samples were collected using the above-mentioned setup. Additionally, we study how well the offline agents are able to learn from an expert policy. To avoid pure imitation, the trajectories gathered by the expert were artificially corrupted with noise, by letting the expert choose a random action with a low probability. Buffers with $100$, $50$, and $25$ samples were collected using using noisy-expert policy for this experiment.

\subsection{Variational Quantum Circuit}

\begin{figure}[tb!]
    \centering
    \includegraphics{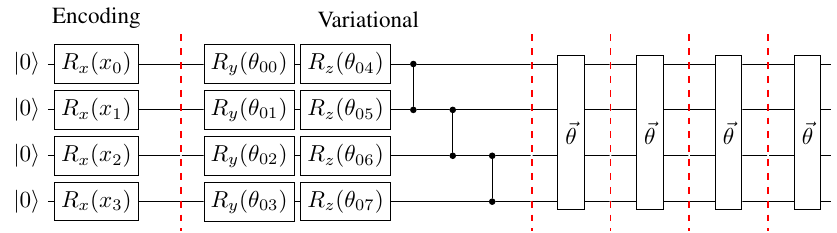}
    \caption{The \gls{VQC} that is used as the function approximator for the \gls{BCQQ} algorithm. \textit{Note}: Each $\vec{\theta}$ block represents the repetition of the variational layer ansatz with different trainable parameters.}
\label{fig:full_vqc}
\end{figure}

The \gls{VQC} used as the function approximator in the \gls{BCQQ} is shown in \cref{fig:full_vqc}. A four-qubit quantum system was chosen as the target system as the CartPole-v1 environment has a four-dimensional state space. Here, each feature of the observation is encoded into the VQC using a single qubit Rx gate on each qubit. The variational block comprises five layers containing four parameterized Ry, and four parameterized Rz gates each. In addition to the parameterized rotational gates, each layer also includes two-qubit CZ entanglement gates with nearest-neighbor connectivity in the circuit layout. We chose the nearest-neighbor connectivity in the circuit layout as this is one of the most commonly available quantum hardware topologies. The CartPole-v1 has an action space of length two. Therefore, the expectation value of the Pauli-$ZZ$ observable on qubits 1 and 2 and Pauli-$ZZ$ observable on qubits 3 and 4 was used to decode the Q-values from the VQC. It is to be noted that the encoding scheme, VQC ansatz, and the decoding schemes used are simple design choices based on previous works \cite{Franz2022, meyer2023quantum}. Different combinations of parameter-shift and SPSA-based gradient estimators along with Adam and AMSGrad optimizers were tested for optimizing the trainable parameters. 

\subsubsection{Data Re-Uploading}
\label{sec:dru}
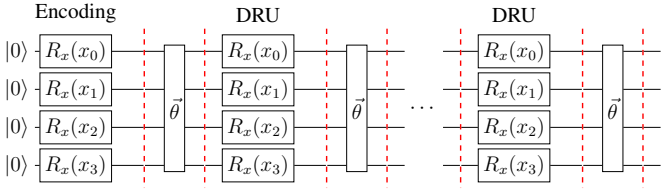
\begin{figure}[htbp!]
\resizebox{\linewidth}{!}{%
  \begin{tikzpicture}[every node/.style={font=\large}]
    \begin{yquant}[operator/separation=1mm, operators/every barrier/.append style={red, thick, shorten <= -4mm, shorten >= -4mm}]
      qubit {$\ket{0}$} q[4];
      [name=encodingL]
      box {$R_x(x_0)$} q[0];
      box {$R_x(x_1)$} q[1];
      box {$R_x(x_2)$} q[2];
      box {$R_x(x_3)$} q[3];
      
      hspace {3mm} -;
      barrier (-);

      [name=variational_start]
      box {$\vec{\theta}$} (q[0, 1, 2, 3]);
      
      [name=variational_end]
      barrier (-);
      [name=DRU_start]
      
      box {$R_x(x_0)$} q[0];
      box {$R_x(x_1)$} q[1];
      box {$R_x(x_2)$} q[2];
      box {$R_x(x_3)$} q[3];
      
      hspace {3mm} -;
      barrier (-);

      box {$\vec{\theta}$} (q[0, 1, 2, 3]);

      barrier (-);
      [draw=none]
      box {$\cdots$} (q[0, 1, 2, 3]);

      barrier (-);
      [name=DRU_last]
      box {$R_x(x_0)$} q[0];
      box {$R_x(x_1)$} q[1];
      box {$R_x(x_2)$} q[2];
      box {$R_x(x_3)$} q[3];
      
      hspace {3mm} -;
      barrier (-);
      
      box {$\vec{\theta}$} (q[0, 1, 2, 3]);

      barrier (-);

    \end{yquant}
    \node[fit=(encodingL), "Encoding"]{};
    \node[fit=(DRU_start), "DRU"]{};
    \node[fit=(DRU_last), "DRU"]{};
    \end{tikzpicture}
    }
    \caption{Quantum agent with standard data re-uploading strategy}
    \label{fig:druVqc}
\end{figure}

Data re-uploading~\cite{Salinas2020} is an encoding strategy where the encoding scheme is repeated throughout the \gls{VQC}. The ordering of the encoding scheme and the trainable layers in a standard VQC is shown in \cref{fig:full_vqc}. Schuld et al.~\cite{Schuld19} show that the more the encoding layer present in the VQC, the larger the frequency spectrum captured by the VQC. Hence, the encoding scheme was re-introduced before every variational layer. \cref{fig:druVqc} represents the circuit generated using the data re-uploading method.

\subsubsection{Cyclic Data Re-Uploading}

\begin{figure}[tb!]
    \centering
    \includegraphics{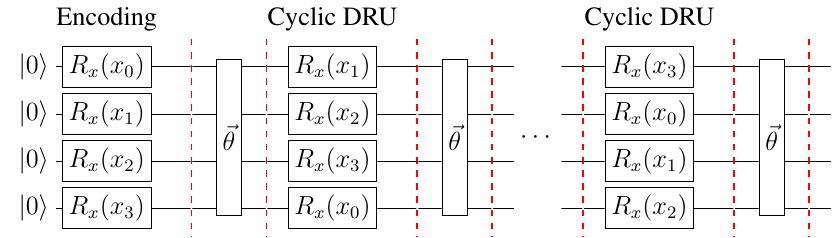}
    \caption{Quantum agent with cyclic data re-uploading strategy}
    \label{fig:CydruVqc}
\end{figure}

It has been established that spreading encoding gates for the feature vector of a given data point throughout the quantum circuit results in a better representation of the data \cite{Periyasamy2022}. Inspired by this, we decided to expose each qubit to all the features of the current input state. We achieve this by slightly modifying the data re-uploading strategy explained in the section on \nameref{sec:dru}. Contrary to the standard approach, we re-introduced the encoding scheme where the input feature vector is shifted one step in a round-robin fashion. Therefore, we call this type of data re-uploading "cyclic data re-uploading". This type of encoding scheme has not been explored in the literature before to the best of our knowledge. \cref{fig:CydruVqc} represents the circuit generated using the cyclic data re-uploading method. 

\subsection{Discrete Batch-Constraint Quantum Q-Learning}

As explained in the sections \nameref{sec:introduction} and \nameref{sec:background}, this work aims to study the advantages gained by using VQCs as function approximators in the discrete BCQ algorithm to learn an optimal policy for solving a given environment. For the same reason, we replaced the generative model \(G_{\omega}\) and the model approximating the optimal policy with two trainable VQCs, as explained in the section \nameref{sec:VQC}. Since we encode data via single qubit rotational gates, each entry of the observation vectors in the dataset is normalized using the encoding scheme presented by \cite{Franz2022}. The overall discrete batch-constraint quantum Q-learning algorithm is summarized in \cref{algo:DiscreteBCQQ}. All the experiments in this study were conducted using the following common hyper-parameters:  discount factor $\gamma = 0.99$, threshold $\tau = 0.3$, and mini-batch size of 32. Three different learning rates $\alpha = [0.01, 0.001, 0.0003]$ were tested for each experiment as the classical and quantum models might need different learning rates for optimal learning on similar problem setups. 

\begin{algorithm}[tbhp!]
\caption{Discrete \gls{BCQQ} training algorithm}
\label{algo:DiscreteBCQQ}
\begin{algorithmic}
    \STATE Normalize dataset \(\mathcal{B}\) between \([-\pi, \pi]\)
    \STATE Initialize encoding unitary \(U(\cdot)\)
    \STATE Initialize Q value approximator using VQC with \(\theta\) and  \(\theta'\)
    \STATE Initialize generative model \(G_\omega\) using VQC with \(\omega\)
    \WHILE{training not converged}
        \STATE Sample mini-batch \(M\) from \(\mathcal{B}\)
        \FORALL{\((s, a, r, s') \in M\)}
            \STATE Get batch-constraint actions \(\tilde{\mathcal{A}}(s')\) from \eqref{eq:BatchConstraintActions}
            \STATE Collect \(a' \in \underset{\tilde{a} \in \tilde{\mathcal{A}}(s')}{\text{argmax}} Q_\theta(|\psi(s')\rangle, \tilde{a})\)
        \ENDFOR
        \STATE Optimize \(\theta\) w.r.t. \(l(\theta)\) 
        \STATE \; \; \;\; \; \; = \(\underset{M}{\mathds{E}}\left(r + \gamma Q_{\theta'}(|\psi(s')\rangle, a') - Q_\theta(|\psi(s)\rangle, a)\right)^2\)
        \STATE Optimize \(\omega\) w.r.t \(l(\omega) = - \underset{M}{\mathds{E}}\text{log}\left(G_\omega (a, |\psi(s)\rangle)\right)\)
        \IF{target VQC update}
            \STATE \(\theta' \leftarrow \theta\)
        \ENDIF
    \ENDWHILE
\end{algorithmic}
\end{algorithm}

\begin{figure*}[tbhp!]
        \centering
        \includegraphics{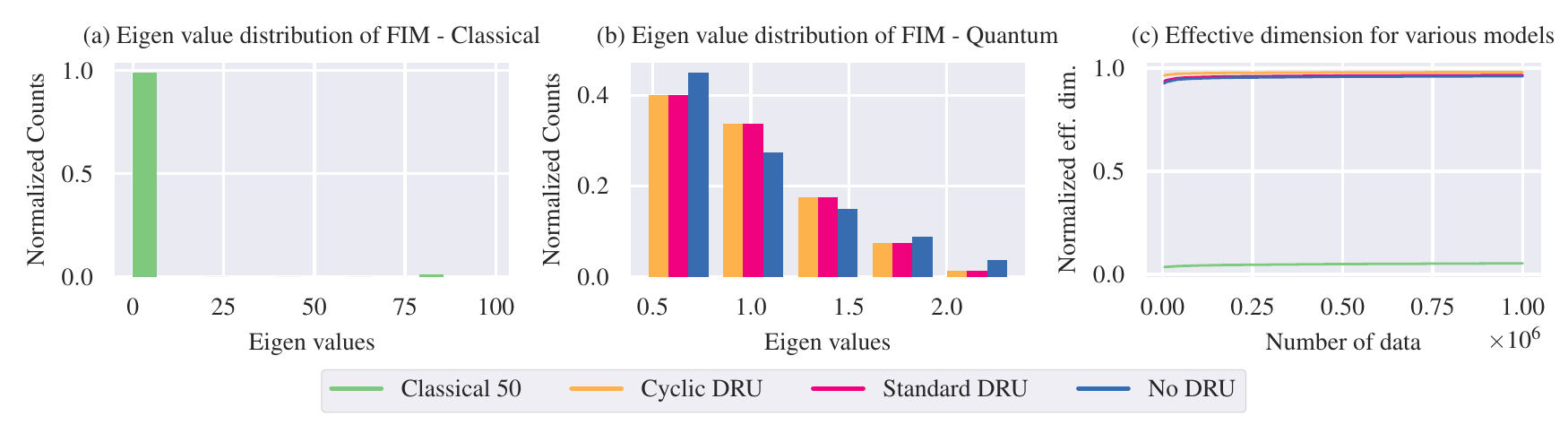}
        \caption{Figure (a) shows the eigenvalue spectrum of average \gls{FIM} for the classical model plotted as a histogram with normalized counts. Figure (b) shows the eigenvalue spectrum of average \gls{FIM} for quantum models plotted as a histogram with normalized counts. Figure (c) shows the effective dimension results for both classical and quantum models. The \gls{FIM} is calculated using 500 data points sampled from the CartPole-v1 states and 100 random parameter sets.}
        \label{fig:effective_dim}
\end{figure*}

\subsection{Model selection}
\label{sec:effective_dimension}

In order to understand the difference between cyclic and standard data re-uploading, we analyze the effective dimension of the resulting models. The effective dimension captures the expressivity of an ML model and is based on the empirical \gls{FIM} \cite{Berezniuk_2020, Abbas2021, Rissanen_1996}. Intuitively, it quantifies the range of different functions a given model can approximate. Furthermore, the eigenvalue spectrum of the \gls{FIM} gives insights into the geometry of the models' parameter space and hence the models' trainability. \cref{fig:effective_dim} shows both the effective dimension and the eigenvalue spectrum of the \gls{FIM} for both quantum and classical models using CartPole-v1 states as input. It becomes apparent that the cyclic data-reuploading strategy results in a slight increase in the effective dimension even though the eigenvalues of the \gls{FIM} are more or less uniformly distributed in all quantum models. Therefore the cyclic data-reuploading strategy was chosen as the quantum model for the experiments. The effective dimension result for the classical model was calculated only for a network with 50 parameters and not for larger networks. This is due to the memory resource bottleneck in calculating the \gls{FIM} for larger networks. Hence, classical networks consisting of higher number of parameters are considered for the experiments. Details regarding the network architectures are explained in the later sections.

\section{Results and Discussion}
\label{sec:results}

\subsection{Training on Random Trajectory}
\label{sec:random_trajectory}

The VQC presented in \cref{fig:CydruVqc} was used as quantum agents for the discrete \gls{BCQQ} algorithm to find an optimal policy, which solves the CartPole-v1 environment from a buffer filled with random environment interactions. \cref{table:dicrete_bcqq} presents the cumulative validation reward averaged over three training runs each, which were performed for a maximum of 25000 gradient update steps. In addition, training was stopped when the agent reached a cumulative reward of 500 in all ten validation environments to reduce the computational cost of the experiments. All experiments were performed on a quantum simulator with Qiskit API~\cite{Qiskit}. SPSA-based gradient estimation along with AMSGrad optimizer and a learning rate of 0.01 resulted in the best average reward by the quantum agent. From \cref{table:dicrete_bcqq}, it can be seen that the quantum agent with cyclic data re-uploading strategy is able to learn an optimal policy for all buffer sizes, even from just 100 random interactions.

\begin{table}[tbhp!]
\huge
\centering
\resizebox{\linewidth}{!}{%
\begin{tabular}{|c|c|c|c|}
\hline
Function Approx.                         & No. of Params. & Buffer Size & Average Reward \\
\hline
\multirowcell{3}{Quantum Agent~\\with \\Cyclic DRU} & \multirow{3}{*}{42x2}          & 1e6         & 500            \\
                                         &                & 1e4         & 500            \\
                                         &                & 1e2         & 500            \\
\hline
\multirowcell{6}{Classical \\ Neural Network     }       & \multirow{3}{*}{67586x2}        & 1e6         & 500            \\
                                         &                & 1e4         & 348.55         \\
                                         &                & 1e2         & 69.33          \\
                                         \cline{2-4}
                                         & \multirow{3}{*}{67x2}           & 1e6         & 342.22         \\
                                         &                & 1e4         & 336.33         \\
                                         &                & 1e2         & 12.88          \\
                                         \hline
\end{tabular}
}
\caption{Average cumulative reward returned by quantum and classical agents. All results are averaged over three training runs}
\label{table:dicrete_bcqq}
\end{table}

\begin{figure*}[thb!]
        \centering
        \includegraphics{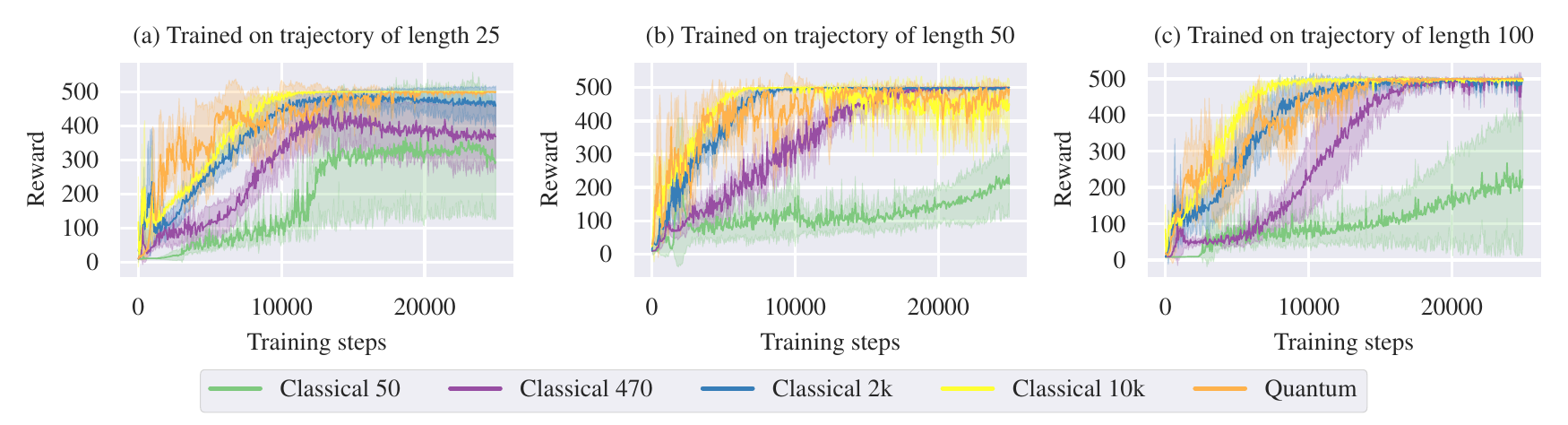}
        \caption{Figure (a), (b), and (c) shows the learning curves of the quantum agent with cyclic data re-uploading strategy and different classical agents trained on partial noisy trajectories of length 25, 50, and 100 respectively. The results shown are averaged over 3 training runs with each evaluation consisting of rewards averaged over 10 random seeds.}
        \label{fig:partial_trajectory_training_results}
\end{figure*}

\subsubsection{Classical Benchmark}

To benchmark the performance of VQCs against classical algorithms, we trained two different classical \glspl{DNN} using the discrete BCQ algorithm on the datasets explained in the section \nameref{sec:dataset}.We chose a fully connected architecture with two hidden layers of 256 or 5 nodes respectively and ReLU activation.
Both networks were trained for a maximum of 100000 steps, instead of 25000 steps as in discrete BCQQ training. This increase was motivated by the fact that these classical \glspl{DNN} can be updated with lower computational cost per update step. However, the early stopping criteria are still used so that the experimental setup remains comparable to the previous one. In the case of the classical network, the Adam optimizer with a learning rate of 0.01 resulted in the highest average reward. The results presented in \cref{table:dicrete_bcqq} show that the classical agent trained with discrete BCQ has difficulty learning an optimal policy from small buffers generated using a random policy. The large NN alone was able to learn an optimal policy in the one million samples case. However, even the larger network was unable to converge to an optimal solution when the size of the training buffer was reduced.

\subsubsection{Globality Testing}

\begin{figure}[h!]
    \centering
    \includegraphics{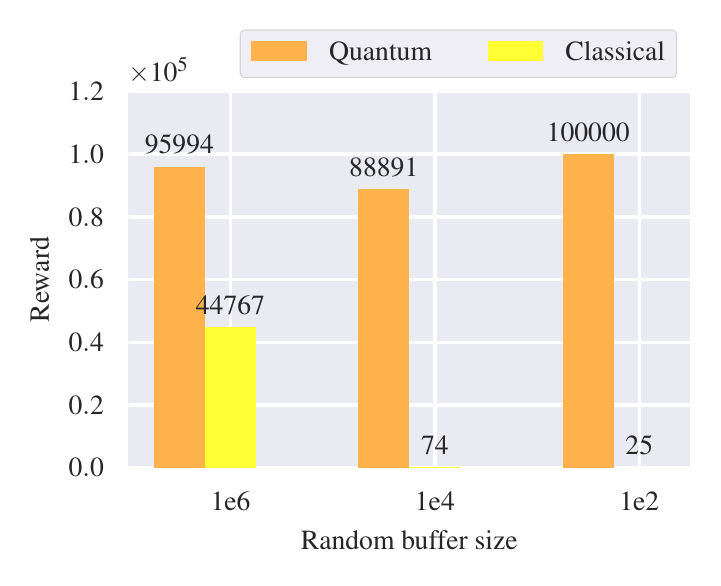}
    \caption{Globality test results in solving CartPole-v1 environment by Quantum and Classical agents. The quantum agent employs cyclic data re-uploading strategy and $42 \times 2$ trainable parameters. The classical agent consists of $67586 \times 2$ trainable parameters }
    \label{fig:100kValidation}
\end{figure}

From \cref{table:dicrete_bcqq} it becomes clear that the VQC with cyclic data re-uploading trained with discrete BCQQ algorithm can learn a policy that attains at least a cumulative reward of 500 for CartPole-v1 environment in all cases. However, it is also interesting to see the maximum cumulative reward beyond 500 that a trained agent can achieve in the CartPole-v1 environment. To this end, we tested the trained agent from the random buffer experiments in the CartPole-v1 environment until either the environment terminated or a cumulative reward of 100000 was achieved. The results of this globality test are shown in \cref{fig:100kValidation}. The results show that the VQC with cyclic data re-uploading not only trains to solve the CartPole-v1 environment, but also learns a more stable policy compared to all other agents.

\subsection{Training on Partial Noisy Trajectory}
\label{subsec:partial_noisy}
Results from the above sections are useful to estimate the capabilities of a VQC with cyclic data re-uploading as a quantum agent. However, in real-world scenarios, the usefulness of offline RL algorithms shines in problems where an RL environment is not available or agent-environment interactions during training are not possible. Here, the agent should be able to learn and hold on to an optimal policy from the clean (or noisy) data collected using an expert policy alone. For this purpose, we collected buffers of 100, 50, and 25 environment transitions, based on a pre-trained expert policy. Here, we compare the performance of a \gls{VQC} with cyclic data re-uploading against 4 classical \glspl{DNN} consisting of 2 hidden layers with 4, 18, 50, and 100 nodes each. Furthermore, to check whether the agent is able to maintain the learned policy, the early stopping condition was not used here. The results from these experiments are presented in \cref{fig:partial_trajectory_training_results}. The quantum agents trained on buffers of size 25 and 100 exhibited a stable learning behavior with a learning rate of 0.001 and when trained with a buffer of size 50 showed a stable behavior with a learning rate of 0.01. Parameter-shift rule with the Adam optimizer was better suited for the quantum agents as the SPSA-based gradients could not hold on to the learned optimal policy. However, all the classical agents showed a stable learning behavior with the same optimizer but a learning rate of 0.0003. All the graphs show that the quantum agent was able to successfully learn to solve the CartPole-v1 environment from partial noisy trajectories. The numerical results indicate that the quantum agent exhibits convergence and performance similar to classical agents with 50 to 250 times more parameters in solving CartPole-v1 environment. The classical agent with a comparable number of parameters struggled in learning an optimal policy when the buffer size was reduced. Our current results indicate a slight advantage depicted by BCQQ in terms of data and the number of trainable parameter requirements. Nonetheless, additional experiments with more complex environments, advanced classical networks, and larger VQCs ( in terms of the number of qubits and the number of parameters), etc. are needed to further substantiate the potential of BCQQ over the BCQ method. However, it is beyond the scope of this work and left for the future.

\subsection{Validation on real quantum hardware}

\begin{figure}[htbp!]
    \centering
    \includegraphics{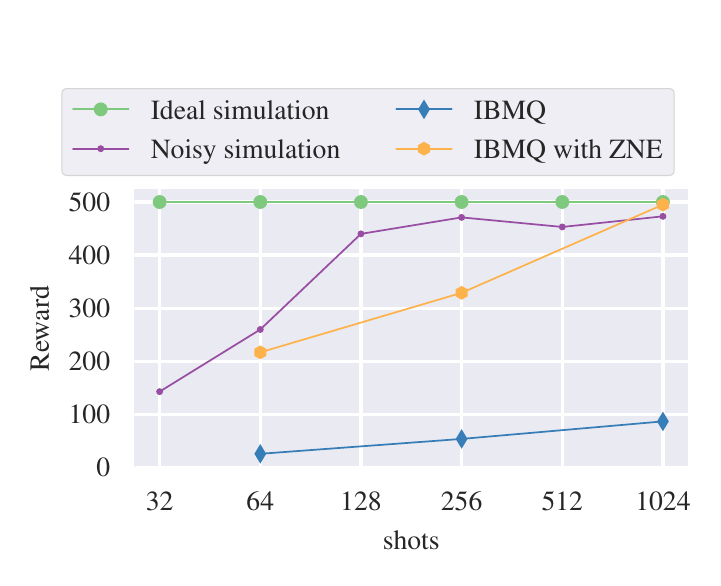}
    \caption{Validation results of a quantum agent trained on an ideal simulator and tested on the noisy simulator, quantum hardware, and quantum hardware with zero noise extrapolation type error mitigation.  }
    \label{fig:ibmq_result}
\end{figure}

All the experimental results presented in the above sections illustrate that the VQC can learn an optimal policy in solving the CartPole-v1 environment with just 100 samples. However, all the experiments presented until now were performed using an ideal quantum simulator that estimates exact expectation values. Whereas the NISQ devices that are currently available are hardware noise prone and only capable of estimating the expectation values. Here, a given quantum circuit is executed multiple times repeatedly to estimate the expectation values. The number of repetitions of the circuit is commonly known as shots. The higher the number of shots, the more accurate the estimated expectation values. To analyze the minimum number of shots required by the agent and the performance of the quantum agent under the influence of hardware noise, we tested the best-performing quantum agent from section \ref{subsec:partial_noisy} on a noisy simulator and the real quantum hardware. The performance of the agent using different numbers of shots is shown in figure \ref{fig:ibmq_result}. 

Validation results with the noisy simulator show that the agent struggles in solving the CartPole-v1 environment using expectation values estimated with 32 or 64 shots. Yet, the agent achieved its peak performance with 128 shots and further increasing the number of shots did not result in a better performance. Here the impact of noise overshadowed the performance gained with the higher number of shots. Further, we also validated the performance of the quantum agent using "ibmq\_mumbai", "ibmq\_kolkata" and "ibm\_algiers" devices. However, due to the higher computational intensity of validation and the scarce availability of quantum resources, we limited the validation to 64, 256, and 1024 shots. The validation results on the IBMQ device show that the agent was only able to achieve a reward of 90 at the most due to the significant impact of noise. Nonetheless, the agent was able to achieve a high reward of 495 with the aid of the zero noise extrapolation error mitigation technique \cite{Temme_2017} and 1024 shots. We speculate that the drop in performance on the IBMQ device could be overcome with further training on the real device. However, further training on the NISQ device is beyond the scope of this work and left for the future.
\section{Conclusion}

In this paper, we propose discrete BCQQ, a batch RL algorithm for discrete action spaces based on VQCs. The key component is a cyclic data re-uploading scheme, where the encoding layers are not only repeated sequentially throughout the circuit, but the order in which the input feature is encoded in which qubit is cyclically shifted from layer to layer. This leads to a slight increase in the model's effective dimension, improving its expressivity without increasing the number of model parameters. Experiments in the CartPole-v1 environment demonstrate that the discrete BCQQ can learn offline from partial trajectories. The results also indicate that the BCQQ algorithm may show better generalization and learning capabilities when learning from smaller data compared to classical networks. For future work, we want to investigate how discrete BCQQ scales with more complex VQCs and evaluate its performance in challenging environments. In addition, we expect that cyclic data re-uploading may lead to improvements in other quantum-based machine learning tasks, such as classification, but leave it for future work.

\FloatBarrier
\section{Acknowledgement}

We thank Dr. Georgios Kontes (Fraunhofer IIS, Nürnberg, Germany), Dr. Steffen Udluft (Siemens AG Technology, Munich, Germany), and Dr. Daniel Hein (Siemens AG Technology, Munich, Germany) for helpful discussions about reinforcement learning. This work was supported by the German Federal Ministry of Education and Research (BMBF), funding program “quantum technologies from basic research to market”, grant number 13N15645. 

\bibliographystyle{IEEEtran}
\bibliography{references}

\end{document}